\documentclass[a4paper,pra,aps,12pt,nofootinbib]{revtex4}

\usepackage[T1]{fontenc}
\usepackage[utf8]{inputenc}
\usepackage{units}
\usepackage{textcomp}
\usepackage{amsmath}    
\usepackage{graphicx}   
\usepackage{verbatim}   
\usepackage{color}      
\usepackage{subfigure}  
\usepackage{hyperref}   
\hypersetup{
citebordercolor=.2 .8 .2,
urlbordercolor=0 .75 1,
backref=true
}
\usepackage{bbm}
\usepackage{nicefrac}
\usepackage{amsfonts}
\usepackage{amsthm}
\usepackage{amssymb}
\usepackage{booktabs}
\usepackage{braket}
\usepackage{fancyhdr}
\usepackage{moreverb}

\makeatletter
\renewcommand*{\email}[1][E-mail: ]{\begingroup\sanitize@url\@email{#1}}
\makeatother
\begin{document}

\title{A new simple class of superpotentials in SUSY Quantum Mechanics}

\author{F. Marques}
\email{fcarmo@fma.if.usp.br}

\author{O. Negrini}
\email{onegrini@fma.if.usp.br}

\author{A. J. da Silva}
\email{ajsilva@fma.if.usp.br}

\affiliation{Institute of Physics, University of S\~{a}o Paulo, São Paulo, SP, Brazil, 05314-970}

\date{November 24, 2011}

\begin{abstract}
In this work we introduce the class of quantum mechanics 
superpotentials \mbox{$W(x)=g\varepsilon(x) x^{2n}$} and study in details the
cases $n=0$ and $1$. The $n=0$ superpotential is shown to lead  
to the known problem of two
supersymmetrically related Dirac delta potentials (well and barrier). 
The $n=1$ case result in the potentials 
\mbox{$V_{\pm}(x)=g^{2}x^{4}\pm2g|x|$}. For $V_{-}$ we present the exact 
ground state solution and study the excited states by a variational technic.
Starting from the ground state of $V_{-}$ and using logarithmic perturbation 
theory we study the ground states of $V_{+}$ and also of $V(x)=g^2 x^4$ and 
compare the result got by this new way with other results for this last potential in 
the literature.
\end{abstract}

\maketitle

\section{Introduction} \label{sec:introducao}

Supersymmetric quantum mechanics (SUSY QM) was first introduced by E.
Witten~\cite{Witten1} \cite{Witten2}, as a simplified model (a $0+1$ dimensional
field theory) to study the possibility of SUSY breaking. Soon it became a research branch in
itself, a way of getting new solutions to problems in quantum
mechanics~\cite{Cooper} \cite{Bagchi} \cite{Junker} \cite{Elso}. Of particular interest, to
our work below, we must cite the many papers in the literature~\cite{BenderLPT3} \cite{CooperLPT}
\cite{BenderLPT1} \cite{BenderLPT2} \cite{SukhatmeLPT} \cite{LeeLPT} \cite{Boya2} \cite{Guilarte}
devoted to the development of technics for treating the anharmonic oscillator $V(x) = \omega^2
x^2 + g^2 x^4$, and other related potentials, which in general do not have exact solutions.

In this work we present a new simple class of superpotentials in SUSY QM, in the form 
\mbox{$W(x)=g\varepsilon(x) x^{2n}$} with $n = 0, 1, 2, \ldots$. The first example of this class, 
i.e., the case $n = 0$, 
was studied long ago in \cite{Boya} and revisited in \cite{Plyushchay1} and \cite{Plyushchay2}. 
One of our results is an analytic solution for the ground state
wave function of the potential $V(x) = g^2 x^4 - 2 g |x|$, an amazing result, 
considering that analytic solutions do not exist for anharmonic 
oscillators. Another result is a new perturbative solution for the ground state of the 
potential $V(x)=g x^{4}$, starting from the solution for the potential 
$V(x) = g^2 x^4 - 2 g |x|$.  Excited states of the potentials 
$V_{\pm}(x)=g^2x^4 \pm 2g|x|$ are also studied by a variational approximation.

The paper is organized as folows. In Sec.~\ref{sec:descricao}, we make a 
brief introduction to the well known case of superpotentials, 
which are monomials of odd powers of $x$, as well as to the SUSY breaking
ones, which are monomials of even powers of $x$. More details of
these solutions can be found in \cite{Cooper} and \cite{FMarques}.
In Sec.~\ref{sec:nossosuperpot}, we study solutions related to the class of 
simple superpotentials
of the form $W(x)=g\varepsilon(x)x^{2n}$ ($n=0,1,2,\ldots$),
where $\varepsilon(x)$ is the sign function. The simple analytical 
solution for the ground state
of the corresponding SUSY system is shown, the already known case $n = 0$ is revised and the case $n = 1$ is studied in more details. The first
one is the ilustrative example of the Dirac delta well and barrier
potentials, which are shown to be SUSY partner potentials associated
with the superpotential $W(x)=g\varepsilon(x)$. The second one,
$W(x)=g\varepsilon(x)x^{2}$, allows us to find an analytical solution
for the ground state of the potential $V(x)=g^{2}x^{4}-2g|x|$. In 
Sec.~\ref{subsec:variacional} and Sec.~\ref{subsec:lpt},
we study the excited states of the potentials $V_{\pm}(x)=g^{2}x^{4}\pm2g|x|$ 
which are derived from 
$W(x)=g\varepsilon(x)x^{2}$. After discussing that exact solutions for the 
excited states cannot be obtained, we apply a variational method 
(Sec.~\ref{subsec:variacional}) to find approximate solutions for 
the energy levels and the wave functions. In 
Sec.~\ref{subsec:lpt}, a new perturbative approach to the ground state of 
the potentials $V(x)=g x^{4}$ and $V(x)=g^2x^4+2g|x|$ is presented. Finally, a 
discussion of the results is presented in the Conclusions.

\section{Our Notation and Definitions on SUSY QM} \label{sec:descricao}

Let us briefly summarize some main concepts in SUSY QM.
For simplicity, we will work in a system of units with the Planck's
constant set as $\hbar=1$ and the particle mass set as $m=1/2$ 
(that is $2m=1$). We start by defining the operators 
$A^{\dagger}$ and $A$:
\begin{equation}
A^{\dagger}=W(x)-ip 
\qquad \text{and} \qquad 
A=W(x)+ip 
\; \text{,}
\label{superescada}
\end{equation}
where $W(x)$ is a given function of $x$ and $p=-id/dx$ is the momentum
operator. From these operators we can construct two hamiltonians:
\begin{equation}
H_{-}=A^{\dagger}A
\qquad \text{and} \qquad 
H_{+}=AA^{\dagger}
\; \text{,}
\label{hamiltfatorizado}
\end{equation}
which in terms of $p$ and $W(x)$ result in: $H_{\pm}=p^{2}+V_{\pm}$. The potentials $V_{\pm}$ are given by the equations ($W^{\prime}\equiv dW/dx$):
\begin{equation}
V_{\pm}(x) = W(x)^{2} \pm W^{\prime}(x)
\; \text{,}
\label{riccati}
\end{equation}
which are Riccati's equations. 

These equations can be understood
in two ways. One way is: given $W(x)$, we can define the hamiltonians
$H_{\pm}$ with potentials $V_{\pm}(x)$. The other is: given the
potential $V_{-}(x)$ (or $V_{+}(x)$), by solving one of the Riccati's
equations, $W(x)$ can be found, the operators $A$ and $A^{\dagger}$
constructed and the partner potential $V_{+}(x)$ (or $V_{-}(x)$)
can be found.

The ground state of a SUSY system is defined as the zero
energy state of $H_{-}$ (this is a choice; changing the function  $W(x) 
\rightarrow -W(x)$ will change the roles of  $H_{-}$ and  $H_{+}$). As $H_{-}=A^{\dagger}A$, its ground state wave function $\psi_{0}^{-}(x)$  can be
got by imposing that it is annihilated by the operator $A$, that
is:
\begin{equation*}
  A\psi_{0}^{-}(x)=\left(W(x)+\frac{d}{dx}\right)\psi_{0}^{-}(x)=0
  \; \text{.}
\end{equation*}
The solution is given by:
\begin{equation}
  \psi_{0}^{-}(x) = \mathcal{N} \exp{\left\{ -\int^{x}W(y)dy \right\} }
\; \text{.}
\label{estfundMQsusi}
\end{equation}

This is a good, physicaly meaningful solution, provided that the 
function~\eqref{estfundMQsusi} is normalizable. Otherwise, a  
zero energy solution does not exist and SUSY is said to be broken.
As it is easy to see, superpotentials obeying the
rule of being positive ($W(x) > 0$) for $x > 0$ and negative ($W(x)<0$)
for $x < 0$ shall manifest SUSY.

Then, starting from $W(x)$, we have two partner hamiltonians, $H_{-}$ and $H_{+}$, one
of them ($H_{-}$, in our choice) having a ground state $\psi_{0}^{-}$
with energy $E_{0}^{-}=0$ and a tower of other states: bound states with energies
$E_{n}^{-}>0$, $n=1,2,3,\ldots$ or scattering states with energies $E^{-} > 0$. The 
hamiltonian $H_{+}$ has bound energy
levels $E_{n-1}^{+}$, $n=1,2,3,\ldots$ with energies related to
the energies of $H_{-}$ by the relation: $E_{n-1}^{+}=E_{n}^{-}$ or scattering energies
$E^{+} > 0$.
Moreover, the eigenfunctions of $H_{-}$ and $H_{+}$ are related according
to:
\begin{align}
 & \psi_{n-1}^{+} = \left(E_{n}^{-}\right)^{-1/2} A \psi_{n}^{-}             \label{conectaestados1} \\
 & \psi_{n}^{-} = \left(E_{n-1}^{+}\right)^{-1/2} A^{\dagger} \psi_{n-1}^{+} \label{conectaestados2}
\; \text{.}
\end{align}

The simplest class of superpotentials manifesting supersymmetry are
monomials of odd power in $x$ , that is:
\begin{equation}
W(x)=gx^{2n+1}
\quad \text{,} \; n=0,1,2,\ldots \; \text{.}
\label{superpotencialpotimpar}
\end{equation}
Using the Riccati equation~\eqref{riccati}, we have for the partner potentials:
\begin{equation}
V_{\pm}(x)=W(x)^{2}\pm W^{\prime}(x)=g^{2}x^{4n+2}\pm g(2n+1)x^{2n}
\; \text{.}
\label{potenciaisparceirospotimpar}
\end{equation}
The ground (normalizable) state of $H_{-}=p^{2}+V_{-}$,
with energy $E_{0}^{-} = 0$ (see Eq.~\eqref{estfundMQsusi}) is given by:
\begin{equation}
  \psi_0^{-}(x)= \mathcal{N} \exp{ \left\{ \frac{-gx^{2n+2}}{(2n+2)} \right\} }
  \; \text{.}
  \label{estfundWpotimpar}
\end{equation}

The first example of a superpotential of the class~\eqref{superpotencialpotimpar}
is $W(x)=gx$. In this case, the associated partner potentials are:
\begin{equation}
V_{\pm}(x)=g^{2}x^{2}\pm g
\; \text{,}
\label{potenciaisparceirospotimparOHS}
\end{equation}
which are simply the potentials of two harmonic oscillators of the
same frequency, with a constant energy shift $g$ added or subtracted.
The ground state of $H_{-}$ have $E_{0}^{-} = 0$. Its excited states
and the states of $H_{+}$ are given by $E_{n}^{-} = E_{n-1}^{+} = 2ng$, 
for $n=1,2,3, \ldots$.  

We will not pursuit the study of this class of superpotentials 
because they are well known. We only mention that the next example of this class,
$W(x)=gx^3$, corresponds to the potentials $V_{\pm}(x)=g^2 x^6{\pm}3gx^2$ 
and their ground state solution is given by \eqref{estfundWpotimpar} with $n=1$.

On the other hand, the class of superpotentials that
are monomials in even powers of $x$, does not give a normalizable
zero energy solution to \eqref{estfundMQsusi} and SUSY is broken.
However, we can introduce the sign function 
$\varepsilon(x)$ and consider superpotentials
of the form \mbox{$W(x)=g\varepsilon(x)x^{2n}$}. For this class of superpotentials, 
a normalizable ground state exists and SUSY is not broken. Thus, in the following we study
this class of superpotentials, specially the $n=0$ and $n=1$ cases.

\section{The class of Superpotentials of the form $W(x)=g\varepsilon(x)x^{2n}$} \label{sec:nossosuperpot}

The case $n=0$ must be treated separately. So, let us consider the
superpotential:
\begin{equation}
W(x)=g\varepsilon(x)
\; \text{,}
\label{superpotEpsilon}
\end{equation}
where $g$ is a positive constant.
For this superpotential \eqref{superpotEpsilon} the Riccati equations 
\eqref{riccati} give the following SUSY partner potentials:
\begin{equation}
V_{\pm}(x)=W(x)^{2}\pm W^{\prime}(x)=\pm 2g\delta(x)+g^{2}
\; \text{,}
\label{riccatiEpsilon}
\end{equation}
where $\delta(x)$ is the Dirac delta function. $V_{-}$ is a delta
well, while $V_{+}$ is a delta barrier, with the energy of the ground state
displaced by $g^{2}$.
The corresponding Schr\"odinger equations are:
\begin{equation}
-\psi^{\pm\prime\prime}(x)\pm2g\delta(x)\psi^{\pm}(x)=\left(E^{\pm}-g^{2}\right)\psi^{\pm}(x)
\; \text{.}
\label{schrEpsilon1rearranja}
\end{equation}
Their solutions are well known \cite{Boya} \cite{Plyushchay1} \cite{Plyushchay2}. The well
($V_{-}$) has a single bound state with energy level $E_{0}^{-}=0$,
binding energy $g^{2}$, and wave function given by:
\begin{equation}
  \psi_{0}^{-}(x) = \sqrt{g} e^{-g |x| }
  \; \text{.}
\end{equation}
All the other eigenstates are plane waves in continuous spectra of
energies, the lowest one starting with $E=g^{2}$. Simple scattering
solutions of the well $V_{-}$ and the barrier $V_{+}$ can be written
as:
\begin{align}
 & \psi_{I}^{\pm}(x)=\mathcal{A}_{\pm}e^{ikx}+\mathcal{B}_{\pm}e^{-ikx}  \quad ,\qquad x \leq 0 \label{solEspalha1}\\
 & \psi_{II}^{\pm}(x)=\mathcal{C}_{\pm}e^{ikx}+\mathcal{D}_{\pm}e^{-ikx} \quad ,\qquad x \geq 0 \label{solEspalha2}
\; \text{,}
\end{align}
where $k=\sqrt{E^{\pm}-g^{2}}$ with $E^{\pm}>g^{2}$ and the respective constants are related according to the bondary conditions $\psi_{II}(0)=\psi_{I}(0)$ and $\psi^{\prime}_{II}(0) = \psi^{\prime}_{I}(0) \pm 2 g \psi(0)$ required by the Dirac delta potential.

Summarizing: the hamiltonian $H_{-}$ has one ground state with energy
$E_{0}^{-}=0$ and continuum of states with energies $E^{-}>g^{2}$
and $H_{+}$ has a continuum of states with $E^{+}>g^{2}$.

To see the role of the supersymmetry in this system, let us consider
a particle crossing the well (or hitting the barrier), coming from $x=-\infty$, such
that we can choose $\mathcal{D}_{\pm}=0$. With the apropriate boundary
conditions through $x=0$, we can determine $\mathcal{B}_{\pm}$ and
$\mathcal{C}_{\pm}$, getting the scattered and the transmited solutions
as functions of the incident amplitudes $\mathcal{A}_{\pm}$. The
results can be written as:
\begin{align}
 & \qquad \psi_{I}^{\pm}(x)=\mathcal{A}_{\pm}\left\{ e^{ikx}+i\frac{\frac{(\mp g)}{k}}{\left(1-i\frac{(\mp g)}{k}\right)}e^{-ikx}\right\}  & ,\qquad x\leq0\label{solEspalhaFinal1}\\
 & \qquad \psi_{II}^{\pm}(x)=\mathcal{A}_{\pm}\frac{1}{\left(1-i\frac{(\mp g)}{k}\right)}e^{ikx} & ,\qquad x\geq0\label{solEspalhaFinal2}
\; \text{.}
\end{align}

It is easy to verify that the solutions $\psi^{-}$ and $\psi^{+}$ are
related by the supersymmetry equations \eqref{conectaestados1} and \eqref{conectaestados2}.
For example, by applying the operator $A$ to $\psi_{I}^{-}(x)$ we
get:
\begin{align*}
 A\psi_{I}^{-}(x)& \propto\left(g\varepsilon(x)+\frac{d}{dx}\right)\left\{ e^{ikx}+i\frac{\frac{g}{k}}{\left(1-i\frac{g}{k}\right)}e^{-ikx}\right\} \\ 
 & \propto\left\{ e^{ikx}+i\frac{\frac{-g}{k}}{\left(1-i\frac{-g}{k}\right)}e^{-ikx}\right\} 
  \propto\psi_{I}^{+}(x) 
  \; \text{,}
\end{align*}
explicitly showing the manifestation of the supersymmetry of the system.

Let us now consider the superpotential:
\begin{equation}
W(x)=g\varepsilon(x)x^{2}
\; \text{,}
\label{superpotEpsilonx2}
\end{equation}
where here also, $g$ is a positive constant. The two partner potentials are given by:
\begin{equation}
V_{\pm}(x) = W(x)^{2} \pm W^{\prime}(x) = g^{2}x^{4} \pm 2g|x|
\; \text{.}
\label{riccatiEpsilonx2}
\end{equation}

In these potentials a term $\delta V=\pm2gx^2\delta(x)$ has been dropped.
The reason is that for the wave functions involved in this problem its action is null. As the potentials $V_{\pm}(x)\rightarrow \infty$ for $x\rightarrow\pm\infty$, the spectra 
of $H_{\pm}=p^2+V_{\pm}$ are discrete and their eigenfunctions are normalizable. If $\delta V$ is treated as a perturbative correction to $H_{\pm}$, its action 
would be non null only if $\int_{-\infty}^{\infty} dx\,x^2 \delta(x) |\psi(x)|^2 \neq0$.
But this condition requires a wave function that near $x=0$ behaves like $f(x)/x$ with $f(0)\neq0$, which is non normalizable and is not in the spectra of $H_{\pm}$.
On the other side, treated as part of $H_{\pm}$, the term $\delta V$ could give non trivial boundary conditions 
for $d\psi/dx$ at $x=0$. To study this possibility we must integrate the Schr\"odinger equation in the interval $x=(-\epsilon,\epsilon)$ for $\epsilon\rightarrow 0$. A non null 
effect of $\delta V$ only comes if
 $\int_{-\epsilon}^{\epsilon}dx\,x^2 \delta(x) \psi(x)\neq0$, which  would require a $\psi(x)$ behaving like $f(x)/x^2$ with $f(0)\neq 0$ that is also, out of the spectra of $H_{\pm}$.

\begin{figure}[h!]
\begin{centering}
\includegraphics[width=0.73\textwidth]{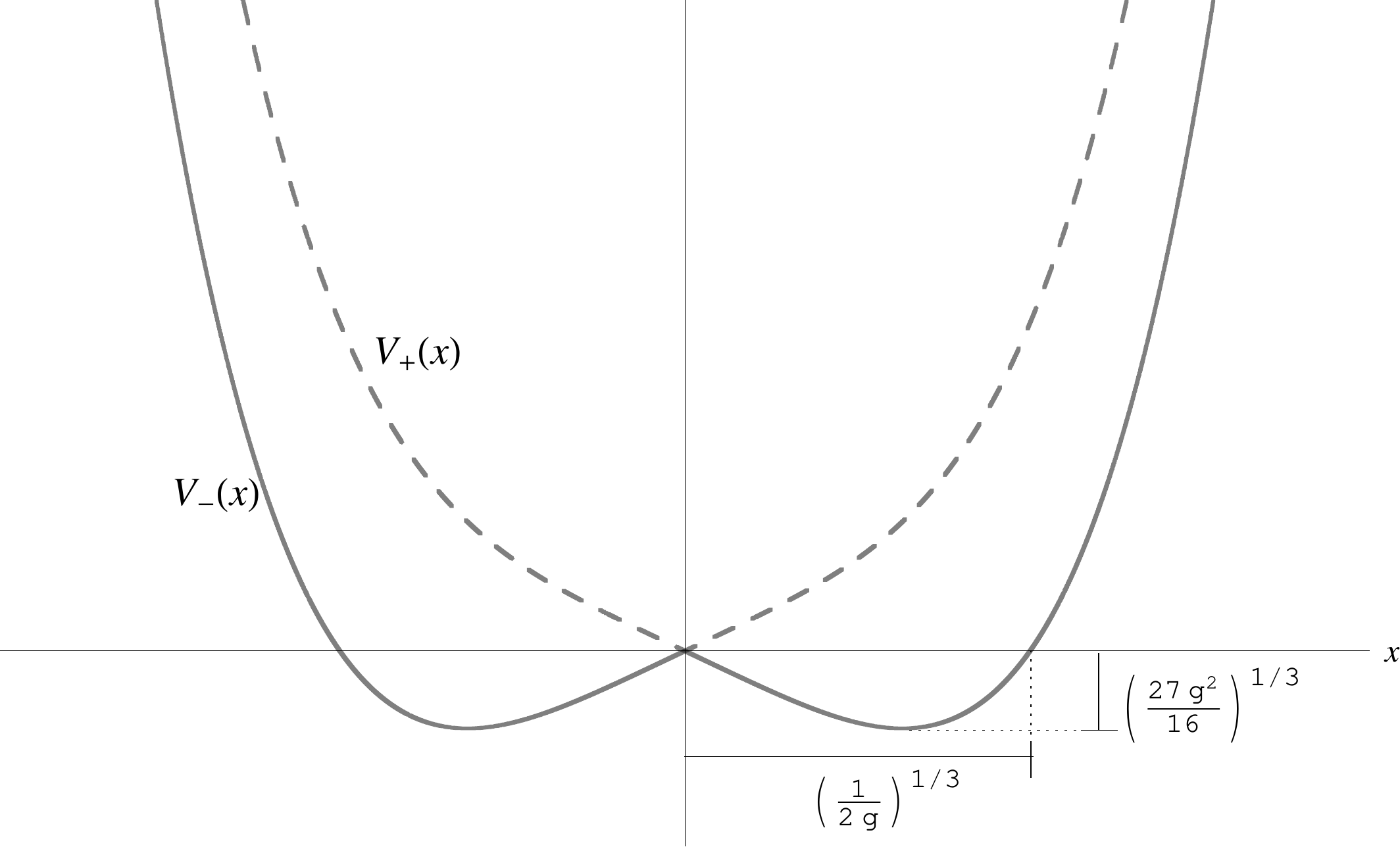} 
\par\end{centering}
\caption{Partner potentials $V_{-}(x)$ and $V_{+}(x)$ associated with the superpotential
$W(x) = g \varepsilon(x) x^2$.}
\label{fig:potEpsilonx2} 
\end{figure}

A representation of these potentials is given in Fig.~\ref{fig:potEpsilonx2}.
As can be seen, $V_{+}$ is a single well potential and $V_{-}$ a
double well potential symmetric in $x$. The corresponding Schr\"odinger
equations read:
\begin{equation}
\left(-\frac{d^2}{dx^{2}}+g^{2}x^{4}\pm2g|x|\right)\psi^{\pm}(x)=E^{\pm}\psi^{\pm}(x)
\; \text{.}
\label{schrEpsilonx2}
\end{equation}

The wave function for the ground state of the double well potential
\mbox{$V_{-}(x)=g^{2}x^{4}-2g|x|$}, has energy $E_{0}^{-}=0$ and is easily obtained
from the equation:
\begin{equation*}
  0=A\psi_0 =\left(g\varepsilon(x)x^{2}+\frac{d}{dx}\right)\psi_0
  \; \text{.}
\end{equation*}
The result (already normalized) is given by:
\begin{equation}
\psi_{0}(x)=\left(\frac{3}{2}\right)^{1/3}\frac{g^{1/6}}{\Gamma\left(1/3\right)^{1/2}}e^{-g|x|^{3}/3}
\; \text{.}
\label{solucaoanaliticaenergiazero}
\end{equation}

This is an interesting result. As it is well known,
exact analytic solutions for the ground (or any excited) state of
the potentials $V(x) = g^{2} x^{4}$ or 
$V(x) = \omega^{2} x^{2} + g^{2} x^{4}$ cannot be obtained. So this exact solution for the
potential $V_{-}$ is somewhat surprising. Another characteristic of
this solution, is that it represents a single lump centered at $x=0$
(which is a local maximum of $V_{-}$) and it is not in the form, as naively expected, of
two lumps centered at the two symmetric minima, $x=\pm(1/2g)^{1/3}$, of 
$V_{-}$ notwithstanding the fact that, in one dimension, any attractive well supports at least a bound state. This happens because the "volume'' of each well is not
big enough to support a bound state (this can be seen in a WKB analysis
of the potential, or even more simply, by the Heisenberg uncertaint
principle. We should only observe that this well size $\Delta x (\Delta E)^{1/2}$ is independent of $g$).

Let us now look for the excited states solutions. Inspired by the
analytic method to solve the one-dimensional simple harmonic oscillator
and by the form of the solution \eqref{solucaoanaliticaenergiazero},
we try a solution of the form
\footnote{In the case of the simple harmonic oscilattor, we supose that the 
solutions are of the form $H(x)e^{-\nicefrac{x^{2}}{2}}$ and, imposing that
those solutions are square integrable, the functions $H(x)$ becomes
restricted to be the Hermite polynomials $\mathcal{H}_{n}(x^{2})$.}:
\begin{equation}
\psi(x)=F(x)e^{-\nicefrac{g|x|^{3}}{3}}
\; \text{.}
\label{assintEpsilonx2}
\end{equation}
Subtituting \eqref{assintEpsilonx2} in the Schr\"odinger 
equation~\eqref{schrEpsilonx2}, it becomes:
\begin{equation}
F^{\prime\prime} - 2 g \varepsilon(x) x^2 F^{\prime}(x) + E F(x)=0
\; \text{.}
\label{schrHeunT}
\end{equation}

For the simple harmonic oscillator, the same steps would lead us to
the Hermite equation. In our case, we get the equation \eqref{schrHeunT},
which is, for a particular choice of parameters, the Triconfluent
Heun equation\cite{Ronv}.

We can go on, look for solutions for the equation \eqref{schrHeunT} through
a power series method. Assuming that $F(x)$ can be written
as:
\begin{equation}
F(x)=\sum_{j=0}^{\infty}a_{j}x^{j}
\label{solucaoHeunTserie}
\end{equation}
and substituting this expression for $F(x)$ in the
differential equation \eqref{schrHeunT}, we find:
\begin{equation*}
  \sum_{j=0}^{\infty}j(j-1)a_{j}x^{j-2}-2g\varepsilon(x)\sum_{j=0}^{\infty}ja_{j}x^{j+1}+E\sum_{j=0}^{\infty}a_{j}x^{j}=0
\; \text{.}
\end{equation*}
Renaming indices and rearranging terms, we have:
\begin{equation*}
  2a_{2}+Ea_{0}+\sum_{j=1}^{\infty}\left[(j+2)(j+1)a_{j+2}-2g\varepsilon(x)(j-1)a_{j-1}+Ea_{j}\right]=0
\; \text{.}
\end{equation*}

Then, given $a_{0}$ e $a_{1}$, this equation is satisfied
if the coefficients $a_{j}$, $j\geq2$, are given by the three terms 
recursion relations:
\begin{align}
 & a_{2}=-\frac{E}{2}a_{0} & ,\qquad j=2\label{coeficientea2} \;\,\, \\
 & a_{j}=\frac{2g\varepsilon(x)(j-3)a_{j-3}-Ea_{j-2}}{j(j-1)} & ,\qquad j\geq3\label{coeficienteaj}
\; \text{.}
\end{align}

The corresponding recursion relation for the the harmonic oscilator potential,
is a simple two terms recursion relation. To get a normalizable solution, we choose
the values of $E$ so as to terminate the series in a polynomial. In this way we get
the set of discretized values of the energy spectrum and the corresponding wave 
functions, that turn up to be the Hermite polynomials (see footnote).

In our case, the recurrence relation \eqref{coeficienteaj}, is a
three terms recurrence relation and there is no way of choosing a subset of values of $E$ to terminate the
series in polynomials, so as to have a normalizable solution. Then, no analytic 
solution can be found and in the next sections we pass to look for 
approximate solutions. In Sec.~\ref{subsec:variacional} a variational approximation is studied 
and in Sec.~\ref{subsec:lpt} a perturbative approximation, that will allows us also, to study 
solutions for the potential $V(x)=g x^4$.

\subsection{Looking for Approximate Solutions by a Variational Method} \label{subsec:variacional}

Let us first apply a variational method. The trial function that we are going 
to use is:
\begin{equation}
\phi(x)=\sum_{j=1}^{m}\alpha_{j}f_{j}(x)
\; \text{,}
\label{functentativa}
\end{equation}
where $j=1,2,\ldots,m.$ and the coefficients $\alpha_{j}\in\mathbbm{C}$
are the variational parameters. The functions $f_{j}(x)$
are chosen to be:
\begin{equation}
f_{j}(x)=x^{j-1}e^{-\nicefrac{g |x|^{3}}{3}}\label{tentativax3}
\; \text{.}
\end{equation}

This trial function corresponds to the previously used in the power series method, 
with the additional restriction of being a finite polynomial of degree $m-1$, 
instead of an infinite series in $x$. 

For the harmonic oscillator, with a very similar choice of the trial function 
we would find exact solutions. In that case,
the variational parameters would be, except for the  normalization, the
coefficients of the Hermite polynomials $\mathcal{H}_{n}(x^{2})$.

Before proceeding, let us consider a convenient change of variables. As can easily be seen, by making the rescaling: $x \rightarrow g^{-1/3}x$ it is possible to 
factor out of the hamiltonians $H_{\pm}$, the constant $g^{2/3}$, that is:
\begin{equation}
H_{\pm}=g^{2/3} \left( -\frac{d^2}{dx^2} + x^{4}\pm2|x| \right)
\; \text{.}
\label{riccatiEpsilonx2reescrita}
\end{equation}

So, in the rest of this section, we will work with $g=1$ and after finding the energy eigenvalues, 
we can restore the dependence of the energy levels in $g$ by multiplying the results by a factor 
of $g^{2/3}$. The restoration of the corresponding wave functions (or trial functions), can also 
be obtained by rescaling $x \rightarrow g^{1/3} x$ in the results.

To go on with the variational method, we construct the expectation value of the energy with 
these trial functions:
\begin{equation}
\begin{aligned}
  E = \frac{ \braket{\phi|H_{\pm}|\phi}}{\braket{\phi | \phi}} 
    = \frac{\sum\limits_{k=1}^{m} \sum\limits_{l=1}^{m}\alpha_k \alpha_l \braket{f_k | H_{\pm} |f_l}}{\sum\limits_{k=1}^{m}\sum\limits_{l=1}^{m}\alpha_k \alpha_l \braket{f_k | f_l }}
\end{aligned}
\end{equation}
and minimize $E$ with respect to the parameters $\alpha _l$. This condition gives the system of linear equations:
\begin{equation} 
\sum_{l=1}^{m}((H_{\pm})_{kl}-E S_{kl}) \alpha_l =0
\; \text{,}
\end{equation}
where we used the notation $H_{kl}=\braket{f_k| H |f_l}$ and $S_{kl}=\braket{f_k|f_l}$.
The values of $E$ that minimize the above system of equations are the eigenvalues of the matrix:
\begin{equation}
M_{kl}=\left(ES_{kl}-(H_{\pm})_{kl}\right)\label{Msim}
\end{equation}
and are obtained by solving the equation $\det{M}=0$. The wave functions corresponding to each of these eigenvalues are got by substituting the value of $E$ in the linear system 
above and solving for the parameters $\alpha_k$. The matrix elements that we need to construct $M_{kl}$ are:
\begin{align}
S_{kl} & =\braket{f_{k}|f_{l}}=\int_{-\infty}^{+\infty}dx\, e^{-\frac{2}{3}|x|^{3}}x^{k+l-2}\label{S}\\
\left(H_{\pm}\right)_{kl} & =\braket{f_{k}|H_{\pm}|f_{l}}=\int_{-\infty}^{+\infty}dx\, e^{-\frac{2}{3}|x|^{3}}\left[-(l-1)(l-2)x^{k+l-4}+2(l\pm1)\varepsilon(x)x^{k+l-1}\right]\label{H}
\; \text{.}
\end{align}
For $(k+l)$ odd, the integrands in \eqref{S} and \eqref{H} are
odd functions and $S_{kl}=\left(H_{\pm}\right)_{kl}=0$. Otherwise,
for $(k+l)$ even, we find:
\begin{align}
S_{kl} & =\left(\frac{3}{2}\right)^{\frac{k+l-4}{3}}\Gamma\left(\frac{k+l-1}{3}\right)\label{Spar}\\
\left(H_{\pm}\right)_{kl} & =-2\left(\frac{3}{2}\right)^{\frac{k+l-3}{3}}\left[\frac{(l-1)(l-2)-(l\pm1)(k+l-3)}{(k+l-3)}\right]\Gamma\left(\frac{k+l}{3}\right)\label{Hpar}
\; \text{.}
\end{align}
With these results, the matrix $M$ \eqref{Msim} gets  the form:

\begin{equation}
M_{\pm}=\begin{pmatrix}(M_{\pm})_{11} & 0 & (M_{\pm})_{13} & 0 &  & \ldots &  & (M_{\pm})_{1m}\\
0 & (M_{\pm})_{22} & 0 & (M_{\pm})_{24} &  & \ldots &  & (M_{\pm})_{2m}\\
(M_{\pm})_{31} & 0 & (M_{\pm})_{33} & 0 &  & \ldots &  & (M_{\pm})_{3m}\\
\\
\vdots & \vdots & \vdots & \vdots &  & \ddots &  & \vdots\\
\\
(M_{\pm})_{m1} & (M_{\pm})_{m2} & (M_{\pm})_{m3} & (M_{\pm})_{m4} &  & \ldots &  & (M_{\pm})_{mm}
\end{pmatrix}
\; \text{.}
\label{Mmatriz}
\end{equation}
In this matrix, all elements in positions $(k,l)$, such that $(k+l)$
is odd are null, while those with $(k+l)$ even are given
by \eqref{Msim} with $S_{kl}$ and $H_{kl}$ respectively given by
\eqref{Spar} and \eqref{Hpar}. To find the energy values we must solve 
the equation: $\det{M}=0$. 

Tables \ref{table:energiasH1Epsilonx2} and \ref{table:energiasH2Epsilonx2}
show some results found for different number ($m$) of parameters and for $g=1$. 
For different values of $g$, the values in the Table must be
multiplied by a factor of $g^{\nicefrac{2}{3}}$, as observed above.

\begin{table}[h!]
\caption{\label{table:energiasH1Epsilonx2} Energy values associated with $H_{-}$ calculated for different numbers of variational parameters.}
\centering
\begin{ruledtabular}
\begin{tabular*}{0.9\textwidth}{@{\extracolsep{\fill}} ccccccccc}
$m$  &   $E^{-}_0$\footnote{For this level, the variational method provides the exact solution.}   &   $E^{-}_{1}$   &   $E^{-}_{2}$   &   $E^{-}_{3}$   & $E^{-}_{4}$  &   $E^{-}_{5}$   &   $E^{-}_{6}$   &   $E^{-}_{7}$  \\
\midrule
 1  &  0.00000 &          &          &          &          &          &          &         \\
 2  &  0.00000 &  2.04441 &          &          &          &          &          &         \\
 3  &  0.00000 &  2.04441 &  5.76541 &          &          &          &          &         \\
 4  &  0.00000 &  1.97852 &  5.76541 & 10.00191 &          &          &          &         \\
 5  &  0.00000 &  1.97852 &  5.54135 & 10.00191 & 14.94174 &          &          &         \\
 6  &  0.00000 &  1.97115 &  5.54135 &  9.49446 & 14.94174 & 20.37028 &          &         \\
 7  &  0.00000 &  1.97115 &  5.51302 &  9.49446 & 14.06558 & 20.37028 & 26.29953 &         \\
 8  &  0.00000 &  1.96991 &  5.51302 &  9.41370 & 14.06558 & 19.02962 & 26.29953 & 32.64399\\
 9  &  0.00000 &  1.96991 &  5.50842 &  9.41370 & 13.90148 & 19.02962 & 24.43194 & 32.64399\\
10  &  0.00000 &  1.96963 &  5.50842 &  9.39868 & 13.90148 & 18.73498 & 24.43194 & 30.18755\\
\end{tabular*}
\end{ruledtabular}
\end{table}

\newpage

\begin{table}[h!]
\caption{Energy values associated with $H_{+}$ calculated for different numbers
of variational parameters.}
\label{table:energiasH2Epsilonx2}
\centering
\begin{ruledtabular}
\begin{tabular*}{0.9\textwidth}{@{\extracolsep{\fill}} ccccccccc}
$m$  & $\,\,\,\,\,\,\,\,\,\,\,\,\,\,\,\,\,\,\,\,$  & $E_{0}^{+}$  & $E_{1}^{+}$  & $E_{2}^{+}$  & $E_{3}^{+}$  & $E_{4}^{+}$  & $E_{5}^{+}$  & $E_{6}^{+}$ \\
\midrule 
1  & $\,\,\,\,\,\,\,\,\,\,\,\,\,\,\,\,\,\,\,\,$  & 2.31447  &  &  &  &  &  & \\
2  & $\,\,\,\,\,\,\,\,\,\,\,\,\,\,\,\,\,\,\,\,$  & 2.31447  & 6.13324  &  &  &  &  & \\
3  & $\,\,\,\,\,\,\,\,\,\,\,\,\,\,\,\,\,\,\,\,$  & 2.04493  & 6.13324  & 10.54940  &  &  &  & \\
4  & $\,\,\,\,\,\,\,\,\,\,\,\,\,\,\,\,\,\,\,\,$  & 2.04493  & 5.63655  & 10.54940  & 15.63469  &  &  & \\
5  & $\,\,\,\,\,\,\,\,\,\,\,\,\,\,\,\,\,\,\,\,$  & 1.99066  & 5.63655  &  9.66470  & 15.63469  & 21.21933  &  & \\
6  & $\,\,\,\,\,\,\,\,\,\,\,\,\,\,\,\,\,\,\,\,$  & 1.99066  & 5.53888  &  9.66470  & 14.30956  & 21.21933  & 27.28556  & \\
7  & $\,\,\,\,\,\,\,\,\,\,\,\,\,\,\,\,\,\,\,\,$  & 1.97666  & 5.53888  &  9.46567  & 14.30956  & 19.36916  & 27.28556  & 33.76558 \\
8  & $\,\,\,\,\,\,\,\,\,\,\,\,\,\,\,\,\,\,\,\,$  & 1.97666  & 5.51611  &  9.46567  & 13.98107  & 19.36916  & 24.86727  & 33.76558 \\
9  & $\,\,\,\,\,\,\,\,\,\,\,\,\,\,\,\,\,\,\,\,$  & 1.97235  & 5.51611  &  9.41524  & 13.98107  & 18.85787  & 24.86727  & 30.72924 \\
10 & $\,\,\,\,\,\,\,\,\,\,\,\,\,\,\,\,\,\,\,\,$  & 1.97235  & 5.51007  &  9.41524  & 13.89369  & 18.85787  & 24.13659  & 30.72924 \\
\end{tabular*}
\end{ruledtabular}
\end{table}

\vspace{1.5cm}

The results in Table 
\ref{table:energiasH1Epsilonx2} and \ref{table:energiasH2Epsilonx2}
reflect the manifestation of SUSY in the system, at least with respect
 to the equality between the energy levels $E_{n}^{-}$
and $E_{n-1}^{+}$, $n>0$, of $H_{-}$ and $H_{+}$. As expected,
the ground state energy of $H_{-}$ is zero and it is not equal to
any energy of $H_{+}$. Moreover, for $n>0$, increasing the
number of variational parameters, we find, mainly for the first levels,
energies $E_{n}^{-}$ more and more closer to $E_{n-1}^{+}$.

Therefore, the better the trial we make, the closer we are
to satisfy the equality between energy levels. Moreover, because the
one parameter trial function for the ground state of $H_{-}$ has
the same form of the exact (analytical) solution, the value $E_{0}^{-}=0$
found is exact and the condition of having a zero energy ground state 
is naturally satisfied.

\begin{figure}[h!]
\begin{centering}
\includegraphics[width=0.7\textwidth]{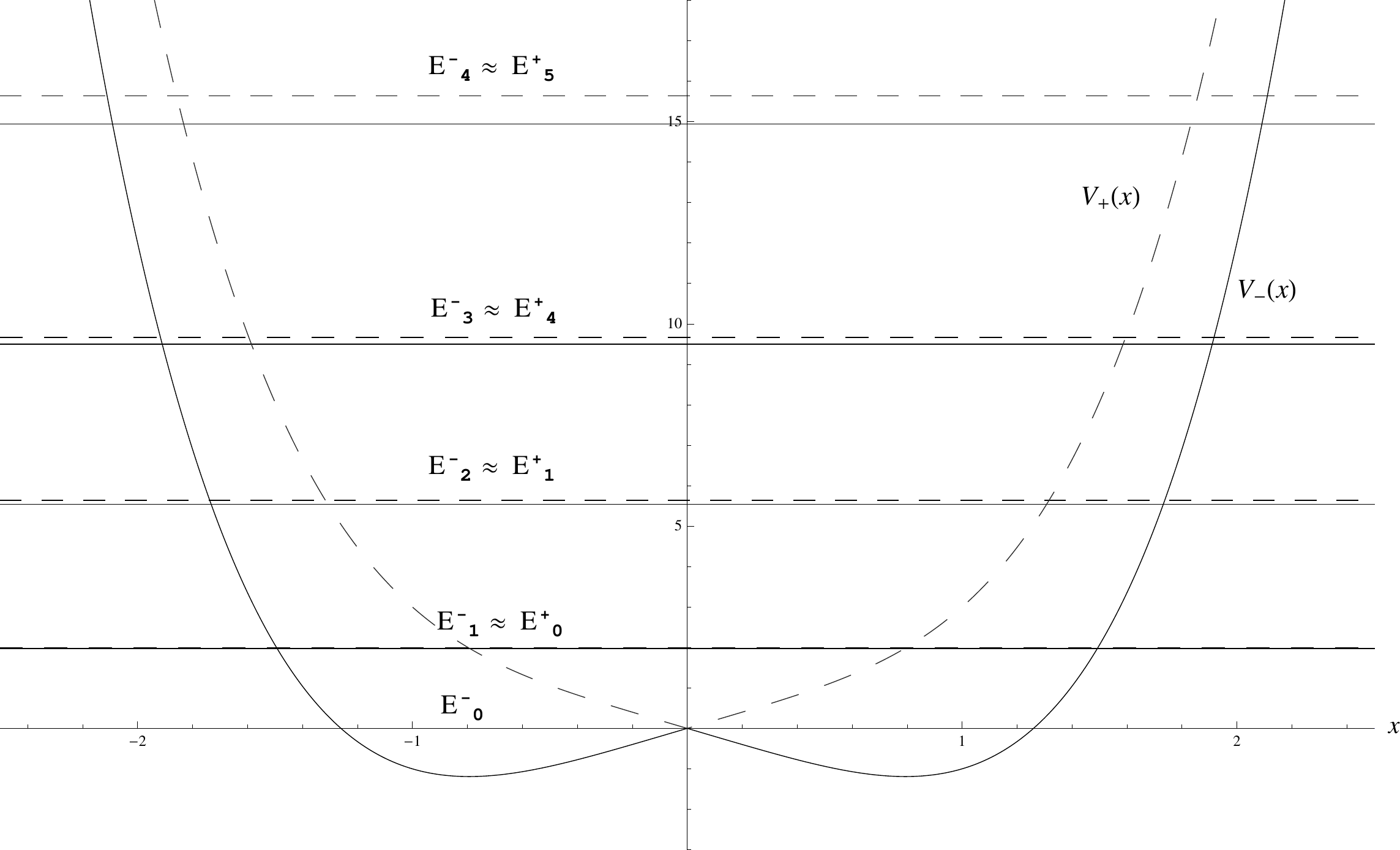} 
\par\end{centering}
\caption{Scheme for the 5 first levels of $H_{-}$ (and 4 first levels of $H_{+}$)
using 6 variational parameters.}
\label{fig:autoenergiasEpsilonx2} 
\end{figure}

Figure \ref{fig:autoenergiasEpsilonx2} shows the first energy
levels of $H_{-}$ and $H_{+}$. We must remember that the values
found are better for increasing number of variational parameters and
for the lowest levels. Thus, for instance, we are supposed to find
for the level $n=4$ a worse approximation than for the level $n=1$.

\newpage

\begin{figure}[h!]
\begin{centering}
\mbox{ \subfigure[Eigenfunctions of $H_{-}$]{\includegraphics[width=0.48\columnwidth]{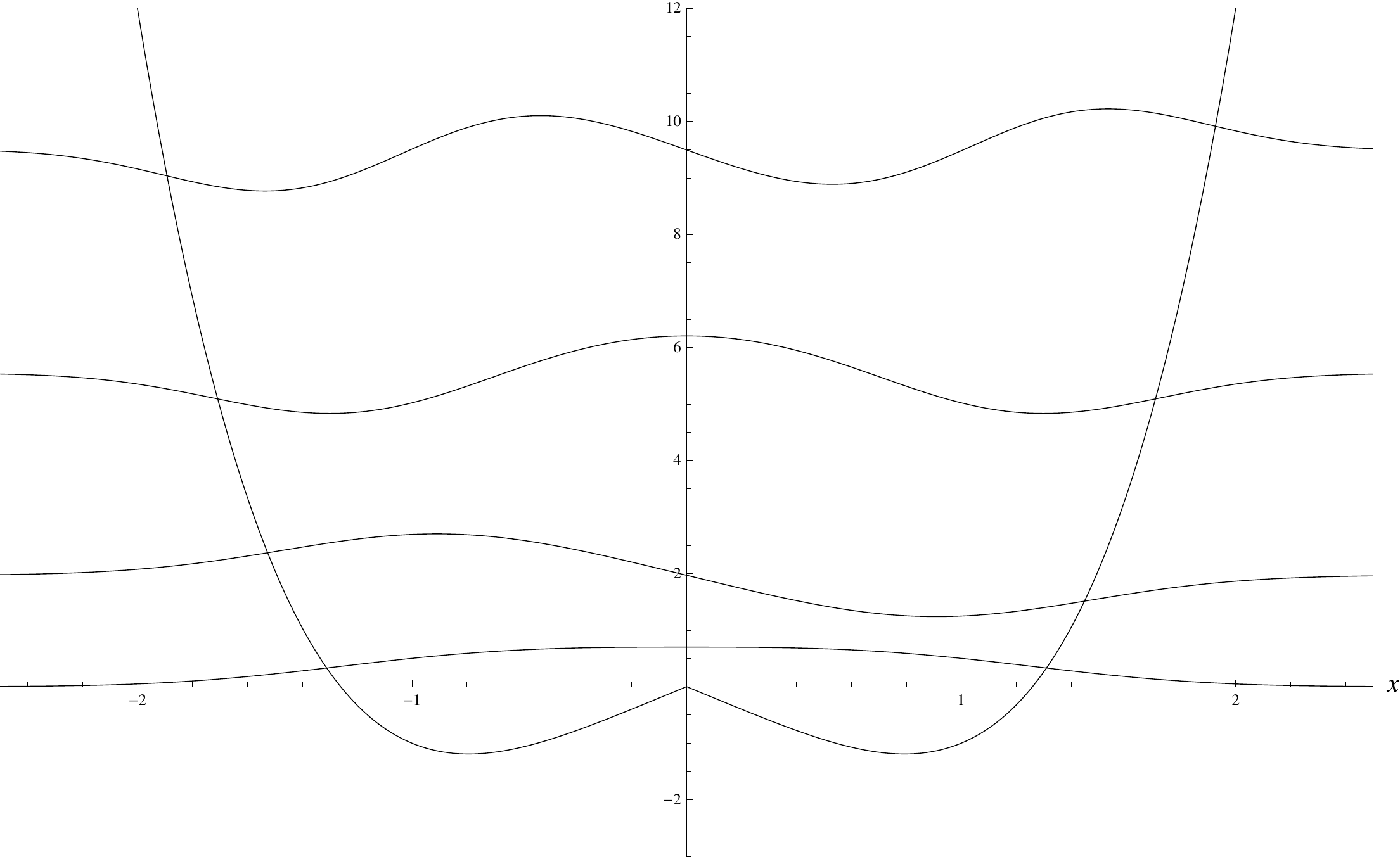}}
\quad{}\subfigure[Eigenfunctions of $H_{+}$]{\includegraphics[width=0.48\columnwidth]{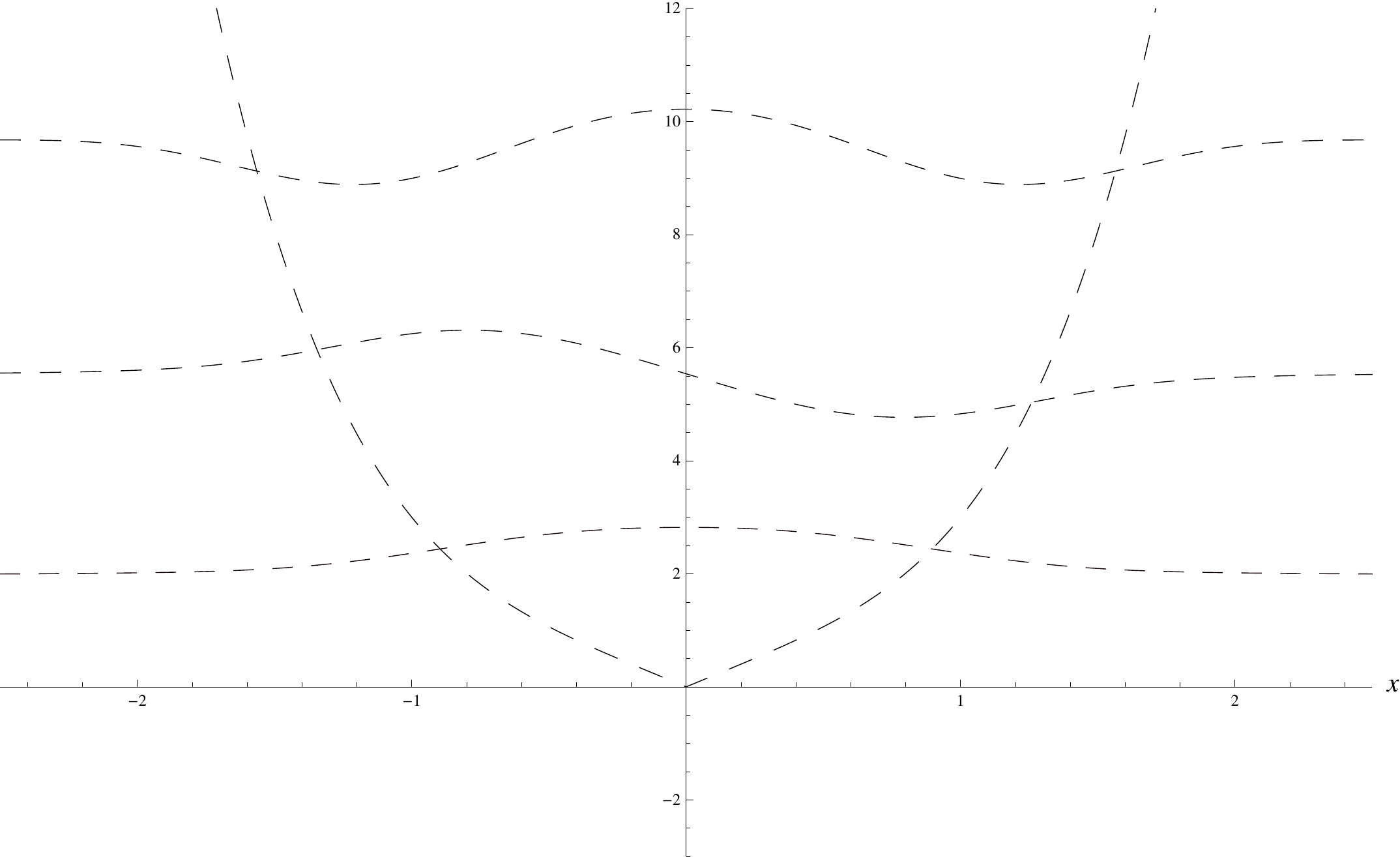}}
} 
\par\end{centering}
\caption{Eigenfunctions of the first levels of $H_{-}$ and $H_{+}$ for 6
variational parameters.}
\label{fig:autofuncoesEpsilonx2} 
\end{figure}

The graphics in Fig.~\ref{fig:autofuncoesEpsilonx2} show the approximations
for the first levels eigenfunctions of $H_{-}$ and $H_{+}$, respectively.
Those approximations were found using 6 variational parameters.

\vspace{1.5cm}

As expected, we see that the eigenfunctions found have well defined
parity, interchanging even and odd solutions with even solutions for
the ground states.

\subsection{Looking for Approximate Solutions by a Logarithmic Perturbation Theory}
\label{subsec:lpt}

We now apply a variant of the logarithmic perturbation theory (LPT) 
to our problem. LPT is explained with more details, for example, in \cite{Cooper},
\cite{BenderLPT3}, \cite{CooperLPT}, \cite{BenderLPT1} or \cite{SukhatmeLPT}.

 Starting from the known solution $\psi_0^{-}$ of $V_{-}$ we can perturbatively obtain the ground state of $V_{+}$, or for example, of the anharmonic potential $V(x)=x^4$. 
We start by writting:
\begin{equation}
V(x; \delta)=V_0(x) + \delta V_1(x)
\; \text{,}
\end{equation}
where:
\begin{align}
& V_0(x)=V_{-}(x) =x^4 - 2|x| \\
& V_1(x)= 4 |x|
\; \text{.}
\end{align}

Observe that $V(x; \delta=1) = V_{+}$ and that $V(x ; \delta=1/2)=x^4$. As we only know the ground 
state of $V_{-}$ we can not go beyond
the first order in the Rayleigh-Schr\"odinger perturbation theory. To bypass this
difficult we will use the so called logarithmic perturbation theory where  only the knowledge $\psi_0^{-}$ is required to calculate the ground state energy level of $V(x; \delta)$ to any order in $\delta$ (at least numerically). For that aim we consider 
the perturbed Schr\"odinger equation:
\begin{equation}
-\Psi" +(V_0+\delta V_1)\Psi=E\Psi
\end{equation} 
and write the expansions:
\begin{align}
& E = E_0+\delta E_1 +\delta^2 E_2+ \ldots \\
& \Psi = \exp{(S_0 +\delta S_1 +\delta^2 S_2 + \ldots)}
\; \text{,}
\end{align}
where $S_1$, $S_2$, etc. are functions and $E_1$, $E_2$, etc. are numbers to be determined. By substituting these
expressions in the Schr\"odinger equation above and equating the terms of same powers in $\delta$ we get the set of equations:
\begin{align}
 S_0^{\prime \prime}+S_0^{\prime \, 2}                                       &=-E_0 + V_2 \\ 
 S_1^{\prime \prime}+2S_0^{\prime} S_1^{\prime}                              &=-E_1 + V_1
 \label{ordem1lptlin} \\
 S_2^{\prime \prime}+2S_0^{\prime} S_2^{\prime} +S_1^{\prime \, 2}           &=-E_2
 \label{ordem2lptlin} \\
 S_3^{\prime \prime}+2S_0^{\prime} S_3^{\prime} +2 S_1^{\prime} S_2^{\prime} &=-E_3 \\
                                                                             & \, \, \, \vdots \nonumber
\end{align}

Starting with $E_0=0$ and $S_0=-|x|^3/3$, 
(that is, $\Psi_{0}(x)=\psi^{-}_{0}=\mathcal{N}e^{-\nicefrac{|x|^{3}}{3}}$), these equations can be 
recursively solved to get $E_k$ and $S_k$ to the desired order in $\delta$. 

Eq.~\eqref{ordem1lptlin} can be rewritten as:
\begin{equation}
  (S'_1 \exp{(2S_0)})^{\prime}=( V_1 - E_1) \exp{(2S_0)}
\; \text{.}
\end{equation}
By substituting $S_0=-|x|^3/3$ and $V_1=4|x|$ in this equation,
integrating both sides in the interval $x=(-\infty,+\infty)$ and observing 
that the integrand of the left side goes exponentially to zero at both 
ends of the integration 
range, we get for $E_1$ the result:
\begin{equation}
\begin{aligned}
E_{1} &=\frac{\braket{\psi_{0}|V_{1}(x)|\psi_{0}}}{\braket{\psi_{0}|\psi_{0}}} 
= \frac{\int_{-\infty}^{+\infty} dx e^{-\frac{2}{3}|x|^{3}} 4|x| }{\int_{-\infty}^{+\infty}dx e^{-\frac{2}{3}|x|^{3}}} \\
&=4 \left( \frac{3}{2} \right)^{ \nicefrac{1}{3} } \frac{ \Gamma(2/3)}{ \Gamma(1/3)} =2.31447
\; \text{.}
\label{ordem1linearcorrecaoenergia}
\end{aligned}
\end{equation}
Inserting this result for $E_1$ back into the same equation and integrating now in the interval $y=(0,x)$ we get:
\begin{equation}
\begin{aligned}
  S'_{1}(x) & =|\psi_{0}(x)|^{-2}\int_{0}^{x}dy |\psi_{0}(y)|^{2}
      \left[E_{1} - V_{1}(y) \right] \\
      & = e^{\frac{2}{3} |x|^{3}} \int_{0}^{x}dy e^{- \frac{2}{3} |y|^{3}}
      \left[ 4 \left( \frac{3}{2} \right)^{\nicefrac{1}{3}} \frac{ \Gamma(2/3)}{ \Gamma(1/3)} - 4|y| \right] \\ 
&= - 2 \left( \frac{2}{3} \right)^{1/3} e^{\frac{2}{3} |x|^{3}} \left[\frac{\Gamma(2/3)}{\Gamma(1/3)} \Gamma(1/3 , 2x^3/3)-\Gamma(2/3 , 2x^3/3) \right]
\; \text{,}
 \label{ordem1linearcorrecaoW}
\end{aligned}
\end{equation}
where $\Gamma(\alpha , x) \equiv \int_x^{\infty} dt \, e^{-t} t^{\alpha - 1}$ are the upper 
incomplete gamma functions\cite{Grad}.

The second order equation \eqref{ordem2lptlin} can also be written in the form:
\begin{equation}
  (S'_2 \exp{(2S_0)})^{\prime}=( -S_1^{\prime \,2}-E_2) \exp{(2S_0)}
\; \text{.}
\end{equation}
Integrating this equation in the interval $x=(-\infty,+\infty)$, and observing that the integrand of the left side goes to zero at both ends of the integration range, we get $E_2$ as an integral over $S'_1$: 
\begin{equation}
 E_2 = - \frac{\braket{\psi_{0}|S'_{1}(x)^2|\psi_{0}}}{\braket{\psi_{0}|\psi_{0}}}
  = - \frac{3}{ \Gamma(1/3)} \left( \frac{2}{3} \right)^{ \nicefrac{1}{3}} 
  \int_{0}^{\infty} dx e^{- \frac{2}{3}|x|^3} S'_1(x)^2
\; \text{.}
  \label{ordem2linearcorrecaoenergia}
\end{equation}
Substituting \eqref{ordem1linearcorrecaoW} in \eqref{ordem2linearcorrecaoenergia}, we find:

\begin{equation}
  E_2 = - \frac{4}{ \Gamma(1/3)} \left( \frac{2}{3} \right)^{ \nicefrac{2}{3} }
  \left\{ \frac{\Gamma(2/3)^2}{\Gamma(1/3)^2} I_{\nicefrac{2}{3}} \left(\tfrac{1}{3},\tfrac{1}{3}\right) 
  + I_{\nicefrac{2}{3}} \left(\tfrac{2}{3},\tfrac{2}{3}\right) 
  - 2 \frac{ \Gamma(2/3)}{ \Gamma(1/3)} I_{\nicefrac{2}{3}} \left(\tfrac{1}{3},\tfrac{2}{3}\right) \right\}
\; \text{,}
  \label{ordem2linearcorrecaoenergiafinal}
\end{equation}
where:
\begin{equation}
  I_{\alpha}(x,y) = \int_{0}^{\infty} dt \, e^t t^{- \alpha} \Gamma(x,t) \Gamma(y,t) \; , \quad
  x > 0 \; , \quad y > 0 \; , \quad 0 < \alpha < 1
\; \text{.}
  \label{integraisdoresultado}
\end{equation}
Evaluating the integrals, the expression \eqref{ordem2linearcorrecaoenergiafinal} gives
$E_2 = - 0.43817$.

Sumarizing: up to the second order, the ground state energy of $V(x;\delta)$ is given by:
\begin{equation}
  E(\delta) = E_0 + \delta E_1 + \delta ^2 E_2
\; \text{,}
  \label{expressaoenergialinear}
\end{equation}
with $E_0=0$, $E_1=2.31447$ and $E_2=-0.43817$. 

For $\delta = 1$, we get for the ground state energy of $V_{+}$, the result: $E_0^{+} = 1.87630$.

For $\delta = 1/2$, we find the result $E_0^{x^4} = 1.04769$ for the ground state energy of the the quartic anharmonic 
potential $V(x)=x^4$. This result can be compared with the exact one given in
\cite{Cooper}, noting that our ``coupling'' constant $g$ is related to their constant 
$\tilde{g}$ by: \mbox{$g^{2/3} = \left( \frac{1}{4} \right)^{1/3} \tilde{g}^{1/3}$}. Thus, multiplying
our result by $\left( \frac{1}{4} \right)^{1/3}$, we find: $\tilde{E}_0^{x^4} = 0.66000$,
differing of \cite{Cooper} only by about $1.2 \%$.

On the other hand, comparing the value of $E_0^{+}$ found here with the most acurate result of
the variational method (see Table \ref{table:energiasH2Epsilonx2}), we see that they differ by about 
$4.9 \%$, what does not seem very good. But, following the suggestion of \cite{CooperLPT} or
\cite{BenderBook}, and substituting the expression \eqref{expressaoenergialinear} by the
corresponding $[1,1]$ Pad\'{e} approximant in $\delta$, we find:
\begin{equation}
  E(\delta) = \frac{E_0 E_1 + (E_1^2 - E_0 E_2) \delta}{E_1 - E_2 \delta}
\; \text{,}
  \label{expressaoenergialinearPade}
\end{equation}
which results (for $\delta = 1$) in $E_0^{+} = 1.94605$. This result now differs from the result of Table 
\ref{table:energiasH2Epsilonx2} only by $1.3 \%$. Doing the same for $\delta = 1/2$ (and
then multiplying by $\left( \frac{1}{4} \right)^{1/3}$), we find $\tilde{E}_0^{x^4} = 0.66597$,
differing from the result of Cooper et all~\cite{Cooper} by only $0.03 \%$. A pretty good result.

\section{Conclusions}

In this paper we studied the class of superpotentials $W(x)=\varepsilon(x) x^{2n}$ in 
SUSY QM. After revisiting the case $n=0$ we went on studying in details the case 
$W(x)=\varepsilon(x) x^2$. As a result we got the exact solution for the ground state 
of the potential $V_{-}(x)=x^4-2|x|$, showed that exact solutions do not exist for the 
excited states and studied these states by a variational method. Finally, starting from
the known ground state of $V(x)=x^{4} - 2|x|$, we obtained the 
ground states for the potentials $V(x)=x^4$ and $V(x)=x^4+2|x|$, by using logarithmic 
perturbation theory. Comparision with other known results in the literature 
and in the paper are given.

Some other approaches and improvements can be used to study this class of superpotentials.
In a forthcoming paper we analyze the solutions for the ground states of  $V_{\pm}(x)=x^{4}\pm2|x|$ by starting with the solutions of $V_{\pm}= x^2\pm C$ and using LPT and the $\delta$ expansion of Bender~\cite{BenderLPT3} and Cooper~\cite{CooperLPT}.

\begin{acknowledgements}
  This work was partially supported by the brazilian agencies Funda\c{c}\~{a}o de Amparo 
  \`{a} Pesquisa do Estado de S\~{a}o Paulo (FAPESP) and Conselho Nacional de 
  Desenvolvimento Cient\'{i}fico e Tecnol\'{o}gico (CNPq). AJS thanks 
  Prof. J. Mateos Guilarte for his warm hospitality at Salamanca University, 
  interesting discussions and for calling his attention for the reference 
  \cite{Boya}. FM and ON thanks Prof. A. Das for useful discussions.
\end{acknowledgements}

\bigskip{\small \smallskip\noindent Updated: \today.}
\end{document}